\documentclass[authoryear,webpdf]{ima-authoring-template}

\usepackage{parskip}
\usepackage{natbib}

\begin{document}

\firstpage{1}

\title[Discussion Network Formation in an Online Professional Development Class]{Discussion Network Formation and Evolution in an Online Professional Development Class: Evidence from a MOOC for K-12 Educators}

\author{Shuhan Ai
\address{\orgdiv{School of Education and Information Studies}, \orgname{University of California Los Angeles}, \orgaddress{\postcode{90095}, \country{USA}}}}

\authormark{S. Ai}

\corresp[]{\href{mailto:aishuhan@ucla.edu}{aishuhan@ucla.edu}}


\abstract{Understanding how educators interact and form peer networks in online professional development contexts has become increasingly important as MOOCs for educators (MOOC-Eds) proliferate. This study examines peer discussion network formation and evolution in ``The Digital Learning Transition in K-12 Schools,'' a MOOC-Ed offered to U.S. and international educators in Spring 2013. Using cross-sectional and temporal exponential random graph models (ERGMs and TERGMs), the study analyzes two network subsamples: the largest connected component (N = 363) and active participants with three or more interactions (N = 227). Results reveal strong reciprocity and transitive closure effects across both networks, with participants six to nine times more likely to reciprocate interactions and over twice as likely to form ties with peers sharing common discussion partners. Assigned discussion group homophily emerged as the strongest predictor of tie formation, while regional homophily and willingness to connect also significantly influenced network structure. Temporal analysis showed discussion activity peaked mid-course before declining sharply, with network structure evolving from broadly distributed participation to concentrated interaction among a tightly connected core. These findings illuminate the mechanisms driving peer-supported learning in online professional development contexts and suggest design implications for fostering sustained educator engagement in MOOC-based learning environments.}


\maketitle

\vspace{1.0cm}

\section{Introduction}

Over the past two decades, massively open online courses (MOOCs) have transformed access to education by enabling learners worldwide to participate in high-quality courses regardless of their geographic location or institutional affiliation \citep{milligan2014supporting}. This transformation has been consequential for professional development (PD) in education, where MOOCs designed specifically for educators, often termed MOOC-Eds, have emerged as scalable platforms for enhancing teaching competencies \citep{koukis2019moocs}. Unlike general-purpose MOOCs, MOOC-Eds are grounded in principles of effective professional learning and structured around four core design principles: multiple voices, self-directed learning, peer-supported learning, and job-connected learning \citep{kellogg2015massively}. Central to this design is the facilitation of peer interaction through discussion spaces organized around professional roles, geographic locations, or shared interests. These peer networks foster communities of practice among geographically dispersed educators and promote communication and reflection across diverse professional contexts. As MOOC-Eds have proliferated in response to growing demand for high-quality online teaching methods \citep{hartshorne2020special}, understanding how educators interact and build connections within online discussion forums has become increasingly important for optimizing course design.

Despite growing interest in network perspectives on online learning, research on peer discussion networks within MOOC-Eds remains limited. Several gaps in the literature warrant attention. First, prior network studies of MOOC-Eds have focused primarily on homophily based on participants' geographic location, professional roles, and teaching experience, while overlooking participants' expressed willingness to connect with peers in online discussion forums \citep{kellogg2014social}. Second, while prior research has applied cross-sectional network analysis to MOOC-Eds, temporal dynamics have received limited attention. Kellogg et al. (2014) examined peer support networks in MOOC-Ed discussion forums using exponential random graph models (ERGMs) and found significant effects of reciprocity and professional role homophily on tie formation. However, their analysis aggregated interactions across the entire course duration, leaving open questions about how these networks form, evolve, and potentially dissolve over time. Moreover, previous studies have primarily examined networks that include all participants regardless of their engagement level \citep{kellogg2014social,wu2021research}. This approach may result in network samples dominated by participants who never replied to a post or replied only once, potentially obscuring interaction patterns among more committed participants. The inclusion of many peripheral or inactive nodes may limit ERGMs' ability to detect meaningful relational processes operating among active participants.

The present study addresses these gaps by extending cross-sectional network analysis of MOOC-Ed discussion forums to incorporate temporal modeling and by examining two network operationalizations: the largest connected component (LC) and the active network (participants with three or more interactions). Using data from ``The Digital Learning Transition in K-12 Schools,'' a MOOC-Ed course offered to U.S. and international educators in Spring 2013 \citep{kellogg2015massively}, this study examines peer discussion network formation and evolution across the course duration. Both cross-sectional ERGMs and temporal ERGMs (TERGMs) are employed to analyze how participant attributes and network structure shape discussion tie formation within and across time periods. This study is guided by three research questions:

\begin{enumerate}
\item What are the demographic and structural characteristics of peer discussion networks in the MOOC-Ed, specifically within the LC and the active network (interactions $\geq$ 3)? How do network structural features change over time as the course progresses?

\item To what extent do participant attributes (e.g., professional role, gender, geographic location, experience level, willingness to connect) and structural mechanisms (e.g., reciprocity, transitivity, popularity) account for tie formation in the aggregated course discussion networks? Do these relationships differ between the LC and the active network?

\item To what extent do participant attributes and network structural mechanisms predict tie formation across successive time periods (e.g., from the first to second quarter)? Do the factors driving network evolution differ across transition periods as the course progresses?
\end{enumerate}

\section{Literature Review}

Recent studies have increasingly applied social network methods to online educational contexts, including examination of how network positions predict learning outcomes, how knowledge diffusion occurs in discussion forums, how peer interactions shape online community structure, and how instructors facilitate online community development \citep{kellogg2014social,kumi2023discussion,ouyang2017influences,wu2021research,wu2021dynamics}. These approaches reveal structural and relational dimensions of online learning that traditional regression methods may overlook, offering insights into the patterns of ties connecting learners within an online learning context.

For instance, Wu and Nian (2021) analyzed a hybrid graduate statistics course using a Facebook learning group over 18 weeks. They found that network centrality measures, particularly outdegree, degree, and closeness centrality, significantly predicted student learning outcomes. Their longitudinal analysis further revealed that the learning community became increasingly interactive over time, with reciprocity rising from 0.638 to 0.667 across three course periods. Similarly, a study of knowledge diffusion in a large Coursera machine learning forum employed exponential random graph models to distinguish between knowledge transfer and knowledge sharing networks, highlighting that users with high degree centrality actively transferred knowledge to others while teaching assistants emerged as particularly influential actors in facilitating forum activity \citep{wu2021research}. Beyond large-scale platforms, Kumi (2023) demonstrated that social network analysis reveals structural and relational dimensions of student interactions that summary statistics and self-reported measures cannot capture. Using graphical representations and centrality measures, this study identified isolated participants, recognized emergent leaders, and exposed both strengths and weaknesses in relational ties within an undergraduate discussion forum.

Within MOOC contexts specifically, work by Kellogg et al. (2014) examined peer networks in two MOOC-Ed courses designed for K-12 educators; their analysis revealed characteristic structural patterns including core-periphery organization, low network density (0.01--0.06), and reciprocity coefficients ranging from 0.15 to 0.26. Building on this work, Zhang et al. (2016) applied simulation investigation for empirical network analysis (SIENA) to examine the dynamic mechanisms driving network change over time in a Chinese MOOC discussion forum. Their study tested four network effects: homophily, reciprocity, transitivity, and preferential attachment. Findings revealed significant positive effects for reciprocity, transitivity, and preferential attachment, yet negative homophily by role, indicating that students preferred interacting with instructors and teaching assistants over fellow peers.

While these studies have employed ERGM and TERGM to examine online discussion forums, the majority focus on MOOCs designed for general student populations and incorporate instructor facilitation within their network analyses. Examining the structural dynamics of peer networks offers important insights for designing effective social learning environments in MOOC-Ed contexts specifically tailored for educator professional development. Understanding who interacts with whom through discussion forum posts, and how these interactions give rise to particular social structures, can illuminate the processes through which social learning occurs among participating educators. The present study therefore explores the mechanisms driving network formation and evolution in a MOOC-Ed context, extending prior work by applying both cross-sectional and temporal exponential random graph models to peer discussion networks among K-12 educators.

\section{Methods}

\subsection{Data}

In Spring 2013, the Friday Institute for Educational Innovation launched the MOOC-Ed Initiative in collaboration with the Alliance for Excellent Education. The initiative began with a six-week pilot course, ``Planning for the Digital Learning Transition in K-12 Schools'' (DLT), designed to help school and district leaders plan and implement K-12 digital learning initiatives \citep{kleiman2013digital}. Regarding tie construction, the initiator of a discussion thread was identified as the default recipient of all responses unless comments were explicitly directed at other participants. Comments directed at non-initiators were identified based on two criteria: (1) use of the forum's ``quote'' feature to reference another participant, or (2) explicit mention of a previous commenter's name. All such comments were visually verified for accuracy \citep{kellogg2015massively}.

The network data derive from two sources. First, the MOOC-Ed registration form collected self-reported demographic information, including participants' professional roles, work settings, years of experience in education, and personal learning goals. Second, the MOOC-Ed discussion forums captured all peer interactions, including posts, comments, and reactions. Database tables containing postings and comments were merged to create a network edge list indicating who interacted with whom, along with participant IDs, timestamps, and discussion content. These interaction data were then linked with registration information to produce a unified dataset containing both peer interaction ties and participant attributes.

To focus specifically on peer-supported learning, posts to or from course facilitators were excluded from the analysis. To examine whether network mechanisms operate differently across participant populations, this study analyzed two subnetworks derived from the overall discussion network (N = 441): (1) LC (N = 363), and (2) the active network (N = 227), defined as participants with a total degree of three or more interactions. Table 1 presents the demographic characteristics and assigned discussion groups of discussants in each network. Participants were predominantly female (approximately two-thirds) and U.S.-based, with technology/media staff comprising the largest professional group (34\%) and teaching experience distributed relatively evenly across experience levels.

\begin{table}[h!]
\centering
\caption{Descriptive Statistics for the Study Node Sample\label{tab1}}
\begin{tabular}{llll}
\toprule
& All OnlinePD & LC Participants & Active Participants \\
& Participants (N = 441) & (N = 363) & (N = 227) \\
\midrule
\multicolumn{4}{l}{\textbf{Region}} \\
International & 32 (7.3\%) & 27 (7.4\%) & 16 (7.0\%) \\
Midwest & 77 (17.5\%) & 63 (17.4\%) & 39 (17.2\%) \\
Northeast & 111 (25.2\%) & 89 (24.5\%) & 58 (25.6\%) \\
South & 169 (38.3\%) & 142 (39.1\%) & 92 (40.5\%) \\
West & 52 (11.8\%) & 42 (11.6\%) & 22 (9.7\%) \\
\multicolumn{4}{l}{\textbf{Country}} \\
Non-US & 32 (7.3\%) & 27 (7.4\%) & 16 (7.0\%) \\
US & 409 (92.7\%) & 336 (92.6\%) & 211 (93.0\%) \\
\multicolumn{4}{l}{\textbf{Gender}} \\
Female & 301 (68.3\%) & 243 (66.9\%) & 150 (66.1\%) \\
Male & 140 (31.7\%) & 120 (33.1\%) & 77 (33.9\%) \\
\multicolumn{4}{l}{\textbf{Educational role}} \\
Teacher & 83 (18.8\%) & 75 (20.7\%) & 48 (21.1\%) \\
Administrator & 91 (20.6\%) & 79 (21.8\%) & 48 (21.1\%) \\
Technology/Media Staff & 162 (36.7\%) & 125 (34.4\%) & 77 (33.9\%) \\
Other Educational Role & 105 (23.8\%) & 84 (23.1\%) & 54 (23.8\%) \\
\multicolumn{4}{l}{\textbf{Grade level involved}} \\
Generalist & 215 (48.8\%) & 168 (46.3\%) & 103 (45.4\%) \\
Primary & 57 (12.9\%) & 50 (13.8\%) & 29 (12.8\%) \\
Secondary & 153 (34.7\%) & 133 (36.6\%) & 88 (38.8\%) \\
Post-Secondary & 16 (3.6\%) & 12 (3.3\%) & 7 (3.1\%) \\
\multicolumn{4}{l}{\textbf{Teaching experience}} \\
$\leq$10 years & 115 (26.1\%) & 100 (27.5\%) & 62 (27.3\%) \\
11--20 years & 150 (34.0\%) & 115 (31.7\%) & 79 (34.8\%) \\
20+ years & 176 (39.9\%) & 148 (40.8\%) & 86 (37.9\%) \\
\multicolumn{4}{l}{\textbf{Expert panelists}} \\
Yes & 20 (4.5\%) & 19 (5.2\%) & 11 (4.8\%) \\
No & 421 (95.5\%) & 344 (94.8\%) & 216 (95.2\%) \\
\multicolumn{4}{l}{\textbf{Connect willingness}} \\
Yes & 69 (15.6\%) & 60 (16.5\%) & 45 (19.8\%) \\
No & 372 (84.4\%) & 303 (83.5\%) & 182 (80.2\%) \\
\multicolumn{4}{l}{\textbf{Assigned discussion group}} \\
AC & 74 (16.8\%) & 68 (18.7\%) & 36 (15.9\%) \\
DL & 50 (11.3\%) & 46 (12.7\%) & 27 (11.9\%) \\
M & 58 (13.2\%) & 43 (11.8\%) & 28 (12.3\%) \\
N & 119 (27.0\%) & 88 (24.2\%) & 60 (26.4\%) \\
PD & 74 (16.8\%) & 60 (16.5\%) & 28 (12.3\%) \\
PS & 66 (15.0\%) & 58 (16.0\%) & 48 (21.1\%) \\
\hline
\end{tabular}
\footnotesize
\begin{flushleft}
Note: LC = Largest Connected Component; Active = Nodes with $\geq$3 total interactions.
\end{flushleft}
\end{table}

The discussion forum was active for 72 days, from April 4, 2013 to June 15, 2013. For temporal analysis, this duration was divided into four quarters: Q1 (days 1--18), Q2 (days 19--36), Q3 (days 37--55), and Q4 (days 56--72). Quarterly subnetworks were constructed to examine how tie formation patterns evolved across the course.

\subsection{Variables}

Both ERGM and TERGM models include network structural variables, a popularity covariate, and homophily covariates. Four structural variables capture endogenous network processes: \emph{edges} represents the baseline propensity for tie formation; \emph{mutual} captures reciprocity, the tendency for ties to be reciprocated; \emph{transitive closure} by assessing whether participants who share common interaction partners are more likely to interact with each other; and \emph{geometrically weighted dyadwise shared partners} (GWDSP) captures two-path connectivity \citep{hunter2006inference,snijders2006new}. Both GWESP and GWDSP use a fixed decay parameter of 0.5 to prevent model degeneracy. For the temporal models, \emph{isolates} is additionally included to capture the tendency for participants to remain disconnected from the discussion network. Moreover, the popularity covariate, \emph{outdegree}, measures whether participants who post more frequently attract more interactions from others. Additionally, homophily covariates assess whether participants preferentially form ties with others who share similar attributes or social identities. These include gender, region, country, professional role, assigned discussion group, grade level, teaching experience, expert status (whether a discussant is a panelist invited to the course), and willingness to connect.

\subsection{Analysis}

\subsubsection{Descriptive Analysis}

This study begins with descriptive statistics to characterize the structure of both the LC and active networks. Metrics include network size, number of edges, density, mean degree (in-degree, out-degree, and total degree), edgewise reciprocity, transitivity, and centralization indices. For longitudinal analysis, these metrics were calculated for each quarterly network (Q1--Q4) to examine how the discussion network structure evolved across the course duration.

\subsubsection{Exponential Random Graph Models (ERGMs)}

ERGMs were used to examine tie formation in the aggregated discussion networks. The likelihood of two nodes forming a tie depend upon attributes of notes and the structure of the network \citep{cranmer2011inferential,leifeld2018temporal}. The ERGM for a network $N$, where $N_{ij} = 1$ if a participant $i$ sent a discussion post to participant $j$, is specified as:

\begin{equation}
P(N, \theta) = \frac{\exp(\theta^\top h(N))}{c(\theta)}
\end{equation}

where $\theta$ is the vector of model coefficients, $h(N)$ is a vector of sufficient statistics, and $c(\theta)$ is a normalizing constant \citep{hunter2008ergm}. In this study, $h(N)$ includes structural statistics, such as reciprocity, GWESP and GWDSP, as well as homophily statistics based on participants characteristics.

\subsubsection{Temporal Exponential Random Graph Models (TERGMs)}

TERGMs extend cross-sectional ERGMs to account for inter-temporal dependency by incorporating parameters that capture how previous network states influence current tie formation \citep{leifeld2018temporal}. The TERGM for network $N^t$ conditional on the previous network is specified as:

\begin{equation}
P(N^t | N^{t-1}, \theta) = \frac{\exp(\theta^\top h(N^t, N^{t-1}))}{c(\theta, N^{t-1})}
\end{equation}

In this study, each quarterly network is modeled conditional on the previous quarter (e.g., $Q2 | Q1$). The joint probability of observing the network series from Q2 to Q4 is:

\begin{equation}
P(N^{Q2}, N^{Q3}, N^{Q4} | N^{Q1}, \theta) = \prod_{t=Q2}^{Q4} P(N^t | N^{t-1}, \theta)
\end{equation}

The TERGM includes the same structural and homophily terms as the cross-sectional ERGM, plus an isolates term capturing participants who remain disconnected in a given quarter.

\subsubsection{Estimation}

Both ERGMs and TERGMs were estimated using maximum pseudolikelihood estimation (MPLE), which approximates the likelihood by replacing the joint probability with the product over conditional dyadic probabilities \citep{strauss1990pseudolikelihood}:

\begin{equation}
\pi_{ij}(\theta) = \text{logit}^{-1} \left( \sum_{r=1}^{R} \theta_r \delta_r^{(ij)}(N) \right)
\end{equation}

where $\pi_{ij}(\theta)$ represents the probability that participant $i$ forms a tie with participant $j$, conditional on all other ties in the network. The term $\delta_r^{(ij)}(N)$ is the change statistic, which measures how much the network statistic $h_r$ changes when a tie between $i$ and $j$ is added. For example, if both participants share the same professional role, the change statistic for role homophily equals 1. A key advantage of MPLE is its computational efficiency: unlike Markov chain Monte Carlo (MCMC) estimation, MPLE does not require simulation, making it well-suited for analyzing large networks.

Cross-sectional ERGMs were fitted using the \texttt{ergm} package \citep{hunter2008ergm}. Temporal ERGMs pooled across all quarters were fitted using the \texttt{btergm} package with bootstrap confidence intervals (R = 100 replications) to address bias in MPLE-based variance estimates \citep{leifeld2018temporal}. To examine whether tie formation mechanisms differed across transition periods, separate formation models were estimated for each consecutive quarter pair (e.g., from Q1 to Q2) using the \texttt{tergm} package with conditional MPLE \citep{krivitsky2014separable}.

\section{Findings}

\subsection{Descriptive Results}

Table 2 presents descriptive statistics for the cross-sectional networks. The LC network comprised 363 participants connected by 1,406 directed ties (density = 0.011), while the active network (N = 227) exhibited higher density (0.024) due to more concentrated interaction among engaged participants. Compared to the LC network, the active network showed higher mean degree, reciprocity, transitivity, and centralization. The relatively higher eigenvector centralization in both networks suggests that while overall participation was distributed, a group of well-connected participants tended to interact with one another. Figure 1 visualizes the cross-sectional discussion networks, with participants grouped by their willingness to connect.

\begin{table}[h!]
\centering
\caption{Descriptive Statistics for Cross-sectional Online Discussion Networks\label{tab2}}
\begin{tabular}{lcc}
\toprule
& LC Network & Active Network ($\geq$3) \\
\midrule
Nodes & 363 & 227 \\
Edges & 1,406 & 1,225 \\
Density & 0.011 & 0.024 \\
Mean In-degree & 3.870 & 5.400 \\
Mean Out-degree & 3.870 & 5.400 \\
Mean Total Degree & 7.750 & 10.790 \\
Reciprocity (Edgewise) & 0.195 & 0.214 \\
Transitivity & 0.126 & 0.138 \\
In-degree Centralization & 0.091 & 0.130 \\
Out-degree Centralization & 0.092 & 0.137 \\
Eigenvector Centralization & 0.280 & 0.269 \\
\hline
\end{tabular}
\end{table}

\begin{figure}[h!]
\centering
\includegraphics[width=1.0\linewidth]{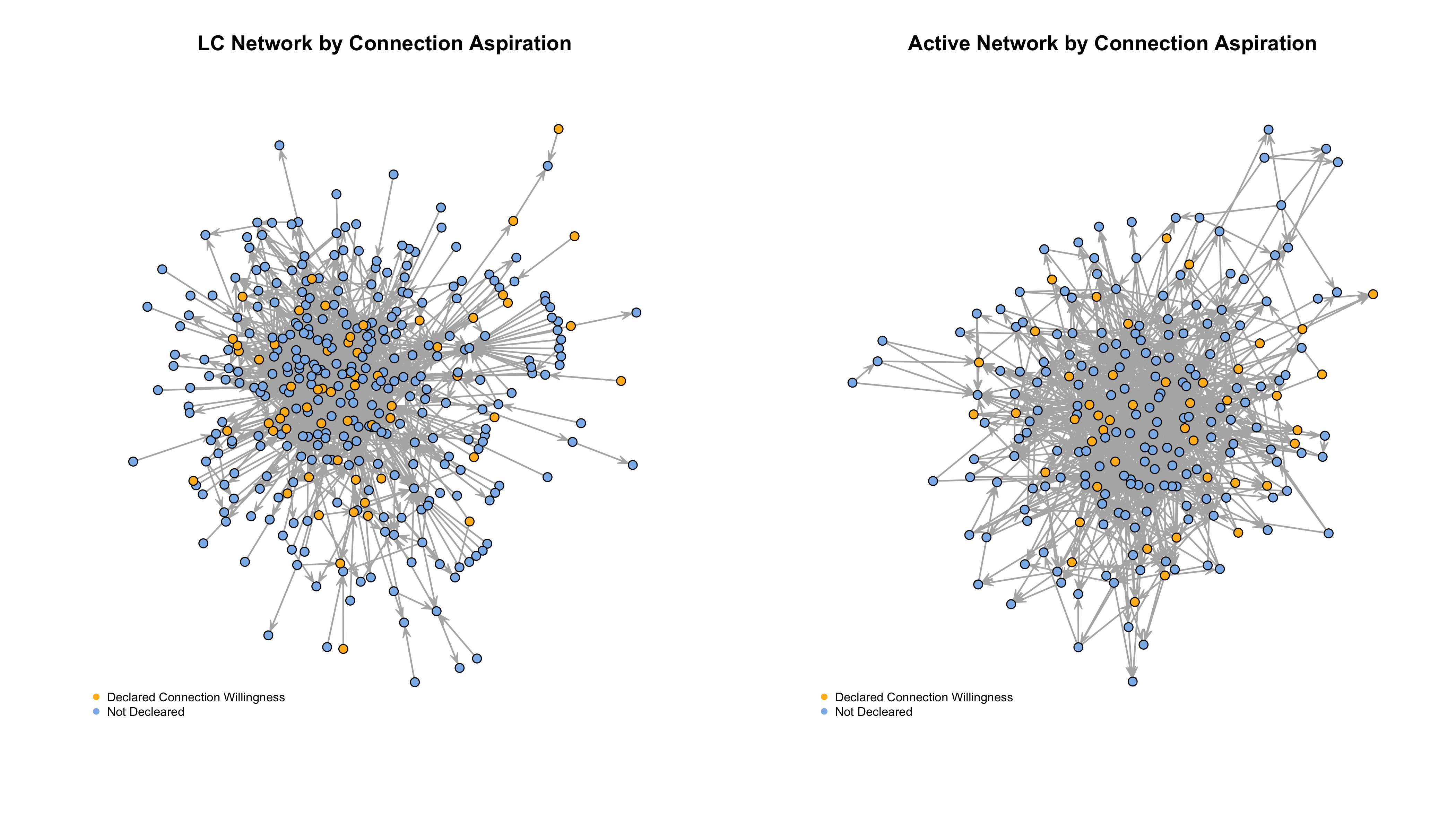} 
\caption{Cross-sectional discussion networks grouped by willingness to connect}
\label{fig1}
\end{figure}

Table 3 and Figure 2 reveal temporal changes in network structure. Discussion activity peaked in Q2 before declining sharply, with edge counts dropping approximately 80\% by Q4. Transitivity increased from Q1 to Q3 in both networks before declining in Q4. Interestingly, while degree and betweenness centralization decreased steadily after Q2, eigenvector centralization increased substantially in Q4. This pattern may suggest that residual activity became concentrated among a small subset of well-connected, influential participants even as overall engagement declined. Figure 3 visualizes how discussion ties evolved across course periods.

\begin{table*}[!t]
\centering
\caption{Descriptive Statistics for Longitudinal Online Discussion Networks by Quarters\label{tab3}}
\begin{tabular}{lcccccccc}
\toprule
& \multicolumn{4}{c}{LC Network} & \multicolumn{4}{c}{Active Network ($\geq$3)} \\
& Q1 & Q2 & Q3 & Q4 & Q1 & Q2 & Q3 & Q4 \\
\midrule
Nodes & 363 & 363 & 363 & 363 & 227 & 227 & 227 & 227 \\
Edges & 515 & 523 & 264 & 104 & 389 & 476 & 260 & 100 \\
Density & 0.004 & 0.004 & 0.002 & 0.001 & 0.008 & 0.009 & 0.005 & 0.002 \\
Mean In-degree & 1.420 & 1.440 & 0.730 & 0.290 & 1.710 & 2.100 & 1.150 & 0.440 \\
Mean Out-degree & 1.420 & 1.440 & 0.730 & 0.290 & 1.710 & 2.100 & 1.150 & 0.440 \\
Mean Total Degree & 2.840 & 2.880 & 1.450 & 0.570 & 3.430 & 4.190 & 2.290 & 0.880 \\
Reciprocity (Edgewise) & 0.093 & 0.191 & 0.197 & 0.019 & 0.113 & 0.202 & 0.192 & 0.000 \\
Transitivity & 0.075 & 0.107 & 0.119 & 0.095 & 0.088 & 0.117 & 0.120 & 0.100 \\
Degree Centralization & 0.042 & 0.039 & 0.034 & 0.017 & 0.055 & 0.060 & 0.053 & 0.025 \\
Betweenness Centralization & 0.047 & 0.030 & 0.007 & 0.002 & 0.066 & 0.063 & 0.016 & 0.004 \\
Eigenvector Centralization & 0.357 & 0.367 & 0.293 & 0.476 & 0.354 & 0.363 & 0.283 & 0.470 \\
\hline
\end{tabular}
\end{table*}

\begin{figure}[h!]
\centering
\includegraphics[width=1.0\linewidth]{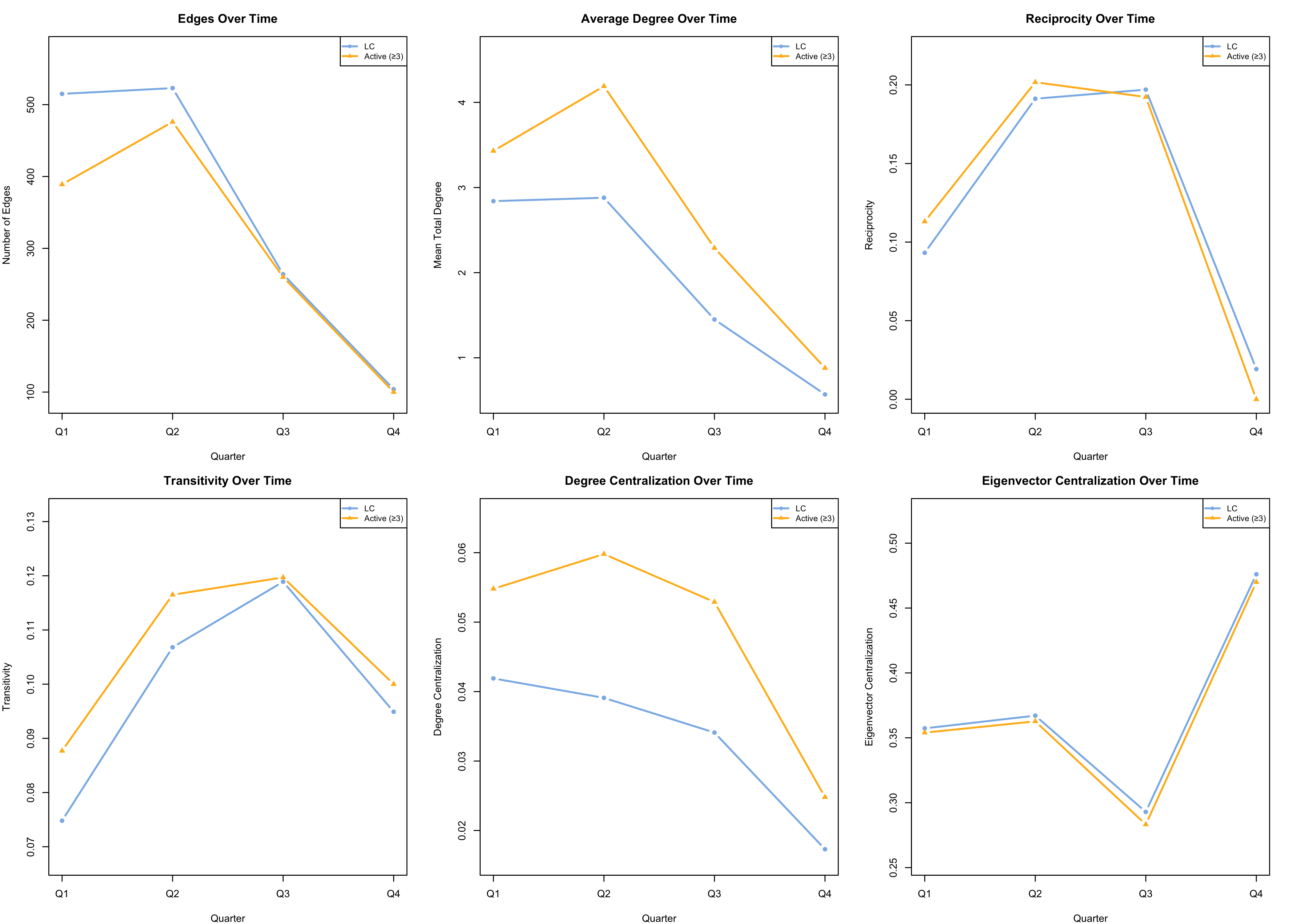}
\caption{Temporal changes in network structure across quarters}
\label{fig2}
\end{figure}

\begin{figure}[h!]
\centering
\includegraphics[width=1.0\linewidth]{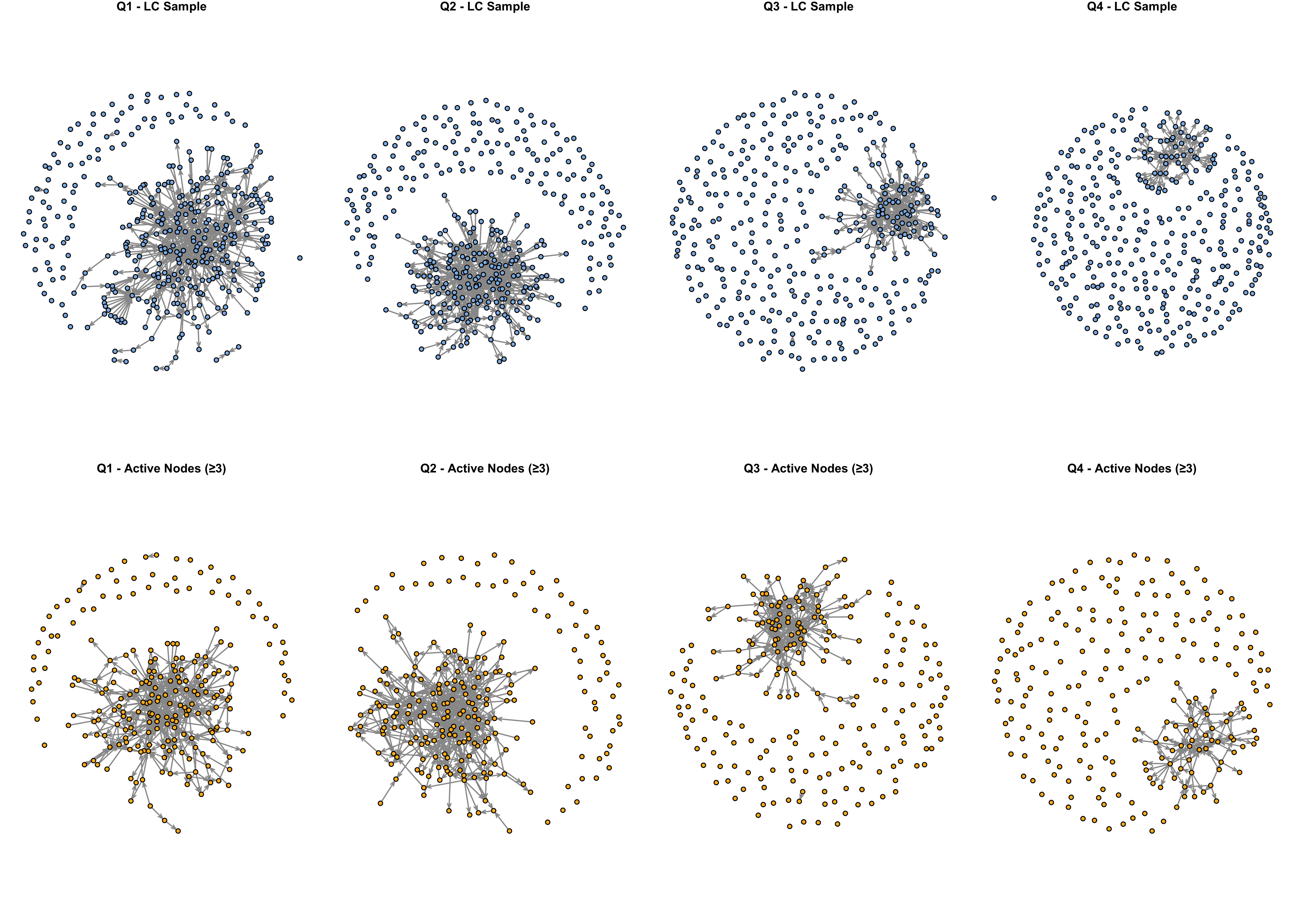}
\caption{Evolution of discussion ties across course periods}
\label{fig3}
\end{figure}

\subsection{Cross-sectional ERGMs Results}

Table 4 presents ERGM results for the LC and active networks. Similar to logistic regression, which predicts a binary outcome from predictor variables, ERGMs predict the presence of a network tie from network structural covariates and node attributes \citep{lusher2013exponential}. Positive coefficients indicate that, controlling for other predictors, corresponding network parameters in the observed network (e.g., ties between participants who both declared willingness to connect) occur more often than expected by chance, and negative coefficients indicate the opposite. The edge term captures baseline tie propensity for tie formation, serving an equivalent role to the intercept in linear regression \citep{handcock2008statnet}.

\begin{table*}[!t]
\centering
\setlength{\tabcolsep}{3pt} 
\footnotesize 
\caption{ERGM Results for Largest Component (LC) Network and Active Network\label{tab4}}
\begin{tabular}{lccclcccc}
\toprule
& \multicolumn{4}{c}{LC ERGM} & \multicolumn{4}{c}{Active ERGM} \\
& b & SE & exp(b) & p-value & b & SE & exp(b) & p-value \\
\midrule
Edge & -5.557 & 0.146 & 0.004 & *** & -4.601 & 0.160 & 0.010 & *** \\
Mutual & 2.031 & 0.091 & 7.623 & *** & 1.843 & 0.092 & 6.313 & *** \\
Transitive Closure & 0.961 & 0.026 & 2.615 & *** & 0.807 & 0.027 & 2.241 & *** \\
Two-Path Connectivity & -0.009 & 0.005 & 0.991 & & -0.037 & 0.006 & 0.964 & *** \\
Outdegree Popularity & 0.011 & 0.005 & 1.011 & * & 0.014 & 0.006 & 1.014 & * \\
Homophily: Role & 0.036 & 0.065 & 1.037 & & 0.070 & 0.070 & 1.072 & \\
Homophily: Group & 0.894 & 0.060 & 2.444 & *** & 0.726 & 0.067 & 2.068 & *** \\
Homophily: Grade level & -0.048 & 0.060 & 0.953 & & -0.069 & 0.065 & 0.933 & \\
Homophily: Gender & 0.058 & 0.059 & 1.060 & & 0.051 & 0.063 & 1.053 & \\
Homophily: Country & -0.083 & 0.086 & 0.921 & & -0.113 & 0.093 & 0.893 & \\
Homophily: Region & & & & & & & & \\
\hspace{3mm}International & 0.793 & 0.251 & 2.211 & ** & 0.877 & 0.289 & 2.403 & ** \\
\hspace{3mm}Midwest & 0.445 & 0.152 & 1.561 & ** & 0.339 & 0.176 & 1.404 & \\
\hspace{3mm}Northeast & 0.307 & 0.110 & 1.359 & ** & 0.261 & 0.118 & 1.298 & * \\
\hspace{3mm}South & 0.141 & 0.077 & 1.152 & & 0.153 & 0.082 & 1.166 & \\
\hspace{3mm}West & 0.080 & 0.230 & 1.083 & & 0.094 & 0.284 & 1.099 & \\
Homophily: Experience & & & & & & & & \\
\hspace{3mm}$\le$10 years & -0.216 & 0.129 & 0.806 & & -0.236 & 0.141 & 0.790 & \\
\hspace{3mm}11-20 years & 0.059 & 0.090 & 1.061 & & -0.019 & 0.095 & 0.981 & \\
\hspace{3mm}20+ years & -0.031 & 0.080 & 0.970 & & 0.073 & 0.087 & 1.075 & \\
Homophily: Expert Status & & & & & & & & \\
\hspace{3mm}Expert & 1.011 & 0.383 & 2.749 & ** & 0.788 & 0.465 & 2.200 & \\
\hspace{3mm}Non-expert & 0.271 & 0.112 & 1.311 & * & 0.270 & 0.123 & 1.309 & * \\
Homophily: Connection Willingness & & & & & & & & \\
\hspace{3mm}Declared & 0.284 & 0.137 & 1.328 & * & 0.292 & 0.136 & 1.339 & * \\
\hspace{3mm}Not declared & -0.179 & 0.064 & 0.836 & ** & -0.149 & 0.069 & 0.862 & * \\
[1.5ex]
Null Pseudo-deviance & 182,167 & & & & 71,370 & & & \\
Residual Pseudo-deviance & 12,471 & & & & 9,876 & & & \\
AIC & 12,515 & & & & 9,920 & & & \\
BIC & 12,730 & & & & 10,115 & & & \\
\hline 
\multicolumn{9}{l}{\footnotesize Note: Significance: *** $p < 0.001$; ** $p < 0.01$; * $p < 0.05$; . $p < 0.1$; Estimation method: MPLE}
\end{tabular}
\end{table*}

Participants showed a strong tendency toward reciprocity (LC: $\exp(b) = 7.623$; active: $\exp(b) = 6.313$, $p < 0.001$), indicating they were six to seven times more likely to respond to a peer from whom they had previously received a response. Participants also exhibited significant transitive closure (LC: $\exp(b) = 2.615$; active: $\exp(b) = 2.241$, $p < 0.001$), suggesting that discussants who shared common interaction partners were over twice as likely to interact with one another. The two-path connectivity parameter was negative and significant only in the active network ($\exp(b) = 0.964$, $p < 0.001$), and participants who posted more frequently attracted slightly more responses from others (LC: $\exp(b) = 1.011$; active: $\exp(b) = 1.014$, $p < 0.05$).

Among homophily effects, assigned discussion group was the strongest predictor: participants in the same group were over twice as likely to interact (LC: $\exp(b) = 2.444$; active: $\exp(b) = 2.068$, $p < 0.001$). Regional homophily was also significant, with international participants showing the strongest within-region preference (LC: $\exp(b) = 0.836$; active: $\exp(b) = 0.862$, $p < 0.01$), followed by participants from the Midwest and Northeast. Expert panelists invited to share their experiences were more likely to interact with fellow experts, though this association was significant only in the LC network ($\exp(b) = 2.749$, $p < 0.01$). Non-expert participants also showed within-group preference in both networks (LC: $\exp(b) = 1.311$; active: $\exp(b) = 1.309$, $p < 0.05$). Additionally, participants who declared a willingness to connect were more likely to form ties with like-minded peers (LC: $\exp(b) = 1.328$; active: $\exp(b) = 1.339$, $p < 0.05$), whereas those who did not express such willingness were less likely to interact with one another (LC: $\exp(b) = 0.836$; active: $\exp(b) = 0.862$, $p < 0.05$). In contrast, professional role, grade level, gender, country, and teaching experience did not significantly predict tie formation. Model fit indices indicated that the active network ERGM demonstrated better fit compared to the LC network ERGM.

\subsection{TERGMs Results}

Table 5 presents the overall TERGM results, which account for temporal dependencies in tie formation across the course duration. Participants demonstrated strong reciprocity (LC: $\exp(b) = 9.550$; active: $\exp(b) = 7.802$) controlling other predictors, indicating they were seven to nine times more likely to respond to peers who had previously responded to them. Transitive closure remained significant (LC: $\exp(b) = 2.585$; active: $\exp(b) = 2.327$, $p < 0.05$). If participant A had discussed with participant B, and participant B had discussed with participant C, participants A and C were over twice as likely to form a discussion tie with each other at the next time point. Moreover, the significant positive isolates parameter (LC: $\exp(b) = 7.529$; active: $\exp(b) = 10.427$, $p < 0.05$) indicates a strong tendency for participants to remain disconnected from the discussion network across time periods.

\begin{table*}[!t]
\centering
\caption{Temporal ERGM Results for Largest Component (LC) Network and Active Network\label{tab5}}
\setlength{\tabcolsep}{2pt} 
\footnotesize
\begin{tabular}{lcccccccc}
\toprule
& \multicolumn{4}{c}{LC Temporal ERGM} & \multicolumn{4}{c}{Active Temporal ERGM} \\
& B & SE & exp(B) & 95\% CI & B & SE & exp(B) & 95\% CI \\
\midrule
Edge & -5.305 & 0.370 & 0.005* & [-6.121; -4.761] & -4.619 & 0.106 & 0.010* & [-4.792; -4.439] \\
Mutual & 2.257 & 0.334 & 9.550* & [1.314; 2.463] & 2.054 & 0.311 & 7.802* & [1.095; 2.342] \\
Transitive Closure & 0.950 & 0.060 & 2.585* & [0.761; 0.997] & 0.845 & 0.066 & 2.327* & [0.643; 0.911] \\
Two-Path Connectivity & -0.012 & 0.019 & 0.988 & [-0.049; 0.002] & -0.041 & 0.018 & 0.960* & [-0.090; -0.013] \\
Isolates & 2.019 & 0.566 & 7.529* & [1.050; 3.324] & 2.344 & 0.185 & 10.427* & [2.034; 2.769] \\
Outdegree Popularity & 0.019 & 0.020 & 1.019 & [-0.004; 0.059] & 0.018 & 0.011 & 1.018* & [0.007; 0.046] \\
Homophily: Role & 0.038 & 0.024 & 1.038* & [0.010; 0.096] & 0.065 & 0.025 & 1.067* & [0.018; 0.119] \\
Homophily: Group & 0.942 & 0.164 & 2.565* & [0.501; 1.180] & 0.795 & 0.180 & 2.214* & [0.258; 0.947] \\
Homophily: Grade level & -0.058 & 0.026 & 0.944* & [-0.116; -0.001] & -0.057 & 0.037 & 0.944 & [-0.132; 0.013] \\
Homophily: Gender & 0.084 & 0.049 & 1.088* & [0.038; 0.225] & 0.071 & 0.049 & 1.073* & [0.006; 0.207] \\
Homophily: Country & -0.149 & 0.083 & 0.862 & [-0.362; 0.008] & -0.160 & 0.098 & 0.852* & [-0.400; -0.032] \\
Homophily: Region & & & & & & & & \\
\hspace{3mm}International & 0.700 & 0.203 & 2.014* & [0.289; 1.235] & 0.720 & 0.227 & 2.055* & [0.229; 1.189] \\
\hspace{3mm}Midwest & 0.375 & 1.488 & 1.455 & [-0.417; 0.642] & 0.294 & 0.210 & 1.342 & [-0.363; 0.525] \\
\hspace{3mm}Northeast & 0.329 & 0.089 & 1.389* & [0.113; 0.523] & 0.314 & 0.063 & 1.369* & [0.163; 0.450] \\
\hspace{3mm}South & 0.199 & 0.109 & 1.220 & [-0.033; 0.423] & 0.199 & 0.120 & 1.221* & [0.002; 0.491] \\
\hspace{3mm}West & 0.020 & 1.310 & 1.020 & [-0.219; 0.198] & 0.027 & 0.183 & 1.027 & [-0.693; 0.096] \\
Homophily: Experience & & & & & & & & \\
\hspace{3mm}$\le$10 years & -0.207 & 0.118 & 0.813 & [-0.351; 0.187] & -0.176 & 0.133 & 0.839 & [-0.386; 0.145] \\
\hspace{3mm}11-20 years & 0.051 & 0.052 & 1.052 & [-0.062; 0.157] & -0.031 & 0.055 & 0.969 & [-0.178; 0.050] \\
\hspace{3mm}20+ years & -0.058 & 0.180 & 0.943 & [-0.641; 0.110] & 0.013 & 0.167 & 1.013 & [-0.465; 0.240] \\
Homophily: Expert Status & & & & & & & & \\
\hspace{3mm}Expert & 0.753 & 5.355 & 2.124 & [-10.831; 0.887] & 0.141 & 0.136 & 1.151* & [0.048; 0.596] \\
\hspace{3mm}Non-expert & 0.141 & 0.164 & 1.151* & [0.102; 0.711] & 0.614 & 5.255 & 1.848 & [-11.263; 1.020] \\
Homophily: Connection Willingness & & & & & & & & \\
\hspace{3mm}Declared & 0.259 & 0.096 & 1.296 & [-0.005; 0.409] & -0.203 & 0.121 & 0.816 & [-0.326; 0.152] \\
\hspace{3mm}Not declared & -0.258 & 0.125 & 0.773 & [-0.380; 0.191] & 0.259 & 0.167 & 1.296 & [-0.186; 0.479] \\
\hline 
\multicolumn{9}{l}{\footnotesize Note: Coefficients with CIs excluding zero are statistically significant at $\alpha = .05$.}
\end{tabular}
\end{table*}

Compared to cross-sectional ERGMs, TERGMs revealed additional homophily effects. Professional role homophily became significant in both networks (LC: $\exp(b) = 1.038$; active: $\exp(b) = 1.067$, $p < 0.05$), indicating that participants were more likely to interact with peers in similar professional roles over time. Gender homophily also emerged as significant (LC: $\exp(b) = 1.088$; active: $\exp(b) = 1.073$, $p < 0.05$), with participants showing preference for same-gender interactions over time. Assigned discussion group remained the strongest homophily predictor (LC: $\exp(b) = 2.565$; active: $\exp(b) = 2.214$, $p < 0.05$). Regional homophily persisted, with international participants (LC: $\exp(b) = 2.014$; active: $\exp(b) = 2.055$, $p < 0.05$) and Northeast participants (LC: $\exp(b) = 1.389$; active: $\exp(b) = 1.369$, $p < 0.05$) showing significant within-region preferences. Notably, country homophily was negative and significant in the active network ($\exp(b) = 0.852$, $p < 0.05$), suggesting that active participants were less likely to interact exclusively with same-country peers over time.

Table 6 presents the separable TERGM formation models, which examine how tie formation processes changed across consecutive quarterly periods. The increasingly negative edge coefficients reflect declining baseline tie formation probability as the course progressed, while the increasingly positive isolates coefficients indicate that participants became less likely to post or reply to discussion posts over time. Reciprocity was strong during Q1 to Q2 and Q2 to Q3 but became non-significant in Q3 to Q4, suggesting that mutual exchange characterized mid-course interactions but dissipated toward course completion. In contrast, transitive closure remained significant throughout all periods. Homophily effects also varied temporally: assigned discussion group homophily was significant during Q1 to Q2 and Q3 to Q4 but not Q2 to Q3, while international participants showed increasing within-region preference, forming ties significantly more often during Q2 to Q3 and Q3 to Q4. Interestingly, participants who had not declared connection willingness were less likely to form ties during Q1 to Q2 but more likely to do so during Q3 to Q4. Unlike the cross-sectional ERGM results, the TERGM findings did not reveal a significant homophily effect for participants who declared a willingness to connect. Additionally, coefficient patterns were largely consistent across the LC and active networks.

\begin{table*}[!t]
\centering
\caption{Separable Temporal ERGM Formation Model\label{tab6}}
\setlength{\tabcolsep}{2pt} 
\scriptsize 
\begin{tabular}{lcccccc}
\toprule
& \multicolumn{3}{c}{LC Temporal ERGM (Formation)} & \multicolumn{3}{c}{Active Temporal ERGM (Formation)} \\
& Q1 to Q2 & Q2 to Q3 & Q3 to Q4 & Q1 to Q2 & Q2 to Q3 & Q3 to Q4 \\
\midrule
Edge & -6.151*** (0.219) & -6.433*** (0.365) & -7.344*** (1.054) & -5.202*** (0.238) & -5.859*** (0.378) & -6.943*** (1.051) \\
Mutual & 2.467*** (0.144) & 2.030*** (0.179) & 0.024 (0.458) & 2.231*** (0.142) & 1.888*** (0.178) & -0.467 (0.543) \\
Transitive Closure (GWESP) & 1.031*** (0.043) & 0.888*** (0.050) & 0.783*** (0.087) & 0.866*** (0.043) & 0.825*** (0.050) & 0.776*** (0.087) \\
Two-Path Connectivity (GWDSP) & -0.018 (0.009) & -0.032* (0.013) & -0.047 (0.030) & -0.059*** (0.011) & -0.039** (0.014) & -0.048 (0.030) \\
Isolates & -0.009 (0.230) & 3.474*** (0.453) & 3.823*** (0.394) & 1.713*** (0.504) & 2.775*** (0.506) & 3.437*** (0.461) \\
Outdegree Popularity & 0.029* (0.013) & 0.126*** (0.024) & 0.012 (0.052) & 0.035* (0.015) & 0.122*** (0.024) & 0.007 (0.053) \\
Homophily: Role & -0.034 (0.106) & -0.081 (0.154) & 0.058 (0.231) & 0.038 (0.110) & -0.060 (0.153) & 0.007 (0.239) \\
Homophily: Group & 0.930*** (0.097) & 0.001 (0.162) & 0.802*** (0.216) & 0.868*** (0.103) & -0.008 (0.164) & 0.756*** (0.223) \\
Homophily: Grade Level & -0.038 (0.097) & -0.172 (0.142) & -0.271 (0.221) & -0.089 (0.102) & -0.174 (0.143) & -0.237 (0.226) \\
Homophily: Gender & 0.083 (0.094) & 0.048 (0.138) & 0.411 (0.220) & 0.082 (0.099) & 0.061 (0.139) & 0.420 (0.225) \\
Homophily: Country & 0.121 (0.152) & -0.293 (0.204) & -0.606* (0.273) & 0.124 (0.160) & -0.393 (0.205) &  -0.603*(0.274)  \\
Homophily: Region & & & & & & \\
\hspace{3mm}International & -0.269 (0.597) & 1.454** (0.551) & 2.161*** (0.647) & -0.425 (0.699) & 1.630** (0.564) & 2.176*** (0.649) \\
\hspace{3mm}Midwest & 0.517* (0.232) & 0.415 (0.452) & -13.723 (424.054) & 0.452 (0.251) & 0.338 (0.446) & -14.333 (552.978) \\
\hspace{3mm}Northeast & 0.544*** (0.157) & -0.006 (0.324) & 1.004** (0.315) & 0.472** (0.165) & 0.028 (0.325) & 0.842* (0.346) \\
\hspace{3mm}South & 0.261* (0.123) & 0.596*** (0.162) & -0.056 (0.293) & 0.204 (0.129) & 0.558*** (0.162) & -0.099 (0.294) \\
\hspace{3mm}West & -0.344 (0.407) & 0.567 (0.599) & -13.636 (501.768) & -0.318 (0.446) & -0.296 (1.017) & -14.006 (772.692) \\
Homophily: Experience & & & & & & \\
\hspace{3mm}$\le$10 years & -0.330 (0.221) & -0.204 (0.310) & 0.492 (0.314) & -0.316 (0.238) & -0.184 (0.310) & 0.436 (0.315) \\
\hspace{3mm}11-20 years & 0.008 (0.147) & 0.021 (0.203) & -0.293 (0.359) & -0.070 (0.149) & -0.006 (0.200) & -0.483 (0.403) \\
\hspace{3mm}20+ years & -0.077 (0.128) & -0.210 (0.194) & -1.392** (0.471) & 0.038 (0.135) & -0.114 (0.195) & -1.351** (0.472) \\
Homophily: Expert Status & & & & & & \\
\hspace{3mm}Expert & 1.303** (0.413) & -10.877 (203.494) & -10.902 (937.070) & 1.050* (0.501) & -10.460 (218.867) & -11.769 (1537.337) \\
\hspace{3mm}Non-expert & -0.210 (0.148) & 0.740* (0.297) & 1.806 (1.015) & -0.046 (0.165) & 0.558 (0.312) & 1.552 (1.018) \\
Homophily: Connection Willingness & & & & & & \\
\hspace{3mm}Declared & 0.026 (0.216) & -0.116 (0.344) & -0.308 (0.771) & 0.176 (0.208) & -0.197 (0.338) & -0.399 (0.773) \\
\hspace{3mm}Not declared & -0.273** (0.103) & -0.068 (0.151) & 0.865** (0.265) & -0.281** (0.109) & 0.009 (0.152) & 0.888*** (0.267) \\
[1.5ex]
Log Likelihood & -3103.28 & -1631.17 & -744.255 & -2474.966 & -1528.12 & -704.742 \\
AIC & 6252.559 & 3308.34 & 1534.509 & 4995.932 & 3102.24 & 1455.484 \\
BIC & 6477.638 & 3533.419 & 1759.588 & 5199.378 & 3305.686 & 1658.93 \\
\hline 
\multicolumn{7}{l}{\footnotesize }
\end{tabular}
\end{table*}

\section{Discussion}

This study empirically examines peer discussion networks in a MOOC-Ed designed for K--12 educator professional development, contributing to a deeper understanding of how online interactions can be structured to foster effective networked learning environments. To investigate the mechanisms underlying network structural dynamics in MOOC-Ed contexts, the study employed both cross-sectional ERGMs and temporal ERGMs. The findings reveal peer interaction patterns that align with well-documented characteristics of online discussion networks, including core--periphery structures, power-law degree distributions, and the predominance of weak ties \citep{kellogg2014social,zhang2016understanding}. Additionally, strong reciprocity and significant transitive closure effects were observed in both LC and active networks, indicating that participants tend to engage in mutual communication and form ties with others who share discussion partners. Notably, these effects remained strong and stable throughout the course duration.

With respect to homophily, educators assigned to the same discussion groups and those sharing regional backgrounds were more likely to communicate with one another, suggesting that structured group assignments can effectively promote interaction and knowledge sharing in online education settings. A particular focus of this study was the role of educators' self-declared willingness to connect in shaping social tie formation. While homophily based on connection willingness was significant in the cross-sectional ERGM, it did not exhibit a consistent effect over time in the temporal models; participants who did not declare such willingness showed mixed effects across different course quarters. Furthermore, discussion activity peaked during the second quarter of the course and declined sharply in the final quarter, whereas eigenvector centralization reached its highest level toward the end of the course. This pattern suggests that as the course progressed, many peripheral participants disengaged or dropped out, leaving a smaller, more densely connected core whose influence became increasingly concentrated. Such dynamics reflect the typical attrition pattern in MOOCs, where participation declines over time and the network evolves from a broadly distributed structure to one dominated by a tightly connected core \citep{wu2021dynamics}.

This study has several limitations. First, the edgelist network construction approach cannot distinguish thread-starters from other participants. Consequently, edges between educators who initiated discussion threads and those who replied directly to them could not be removed. This introduces asymmetry in network interpretation and potentially confounds the analysis of peer interaction patterns. Thread-starters occupy a structurally privileged position similar to facilitators---they set the discussion agenda and naturally receive more replies simply by virtue of posting first. This creates artificial inflation of their in-degree centrality that reflects thread-starter status rather than genuine peer influence or recognition within the learning community. Second, this study analyzed a single MOOC-Ed course offering, which limits generalizability across several dimensions. The findings may not extend to MOOCs with different pedagogical designs, platform features, or instructor facilitation approaches. Additionally, generalizability may be limited to courses serving similar populations (i.e., K-12 educators) or addressing similar content domains (i.e., digital learning transition). Lastly, this study examined only how peer discussion networks form and evolve during the course period, leaving the factors that predict network dissolution or persistence unexplored. Future study could illuminate how online discussion ties persist or dissolve over time.

\clearpage %

\bibliographystyle{plainnat}

\end{document}